\documentclass[10pt]{article}
\usepackage{amsmath,amssymb}
\usepackage{hyperref}
\usepackage{units}
\usepackage{color}
\usepackage[T1]{fontenc}
\usepackage[utf8]{inputenc}
\usepackage{authblk}
\usepackage{bm}
\renewcommand{\vec}[1]{\bm{#1}}

\title{Exact diagonalization of the $d-$dimensional confined quantum harmonic oscillator
\thanks{%
PACS:03.65.Ge} 
}
\author[]{Kunle Adegoke\thanks{Corresponding author: adegoke00@gmail.com}}
\author[]{Adenike Olatinwo}
\author[]{Henry Otobrise}
\author[]{\mbox{Funmi Akintujoye}}
\author[]{Afees Tiamiyu}
\affil{Department of Physics and Engineering Physics, \mbox{Obafemi Awolowo University}, Ile-Ife, Nigeria}

\begin{document}

\date{}

\maketitle

\begin{abstract}
\noindent In the existing literature various numerical techniques have been developed to quantize the confined harmonic oscillator in higher dimensions. In obtaining the energy eigenvalues, such methods often involve indirect approaches such as searching for the roots of hypergeometric functions or numerically solving a differential equation. In this paper, however, we derive an explicit matrix representation for the Hamiltonian of a confined quantum harmonic oscillator in higher dimensions, thus facilitating direct diagonalization.

\end{abstract}

\tableofcontents

\section{Introduction}

The $d-$dimensional confined harmonic oscillator (cho) of mass $m$ and frequency $\omega$ is described by the Hamiltonian
\[
H(\vec x)=-\frac{\hbar^2}{2m}{\vec\nabla}^2+\frac 12m\omega^2{\vec x}^2\,,
\]
where $\vec x=(x_1,x_2,\ldots,x_d)$ with $|x_i|\le L$, ${\vec x}^2={\vec x}^t{\vec x}$ and $\vec\nabla$ is the \mbox{$d-$dimensional} cartesian gradient operator. $H(\vec x)$ being a Kronecker sum, we can also write
\[
H(\vec x)=\sum_{i=1}^d H_i(x_i)\,,
\]

where
\[
\begin{split}
H_i(x_i)=-\frac{\hbar^2}{2m}\frac{\partial^2}{\partial x_i^2}+\frac{1}{2}m\omega^2x_i^2,\quad |x_i|\le L\,.
\end{split}
\]

We also note that $H(\vec x)=T(\vec x)+V(\vec x)$, where
\[
T(\vec x)= \sum_{i = 1}^d {T_i (x_i )} \mbox{ and } V(\vec x) = \sum_{i = 1}^d {V_i (x_i )}\,, 
\]
with
\[
T_i(x_i)=-\frac{\hbar^2}{2m}\frac{\partial^2}{\partial x_i^2} \mbox{ and } V_i(x_i)=\frac{1}{2}m\omega^2x_i^2,\quad (i=1,2,\ldots,d)\,.
\]

Various techniques have been employed by researchers to numerically diagonalize the Hamiltonian of a confined oscillator. These methods usually involve searching for roots of hypergeometric functions, as can be seen for example in references~\cite{jaber} and~\cite{montgomery}. In~\cite{campoy} a method based on the expansion of the wavefunction as well as numerical integration of an ordinary differential equation were used to obtain the energy eigenvalues and wavefunctions of a one-dimensional confined oscillator. 

\bigskip

In this paper we will derive an explicit matrix representation for the Hamiltonian of the confined \mbox{$d-$dimensional} harmonic oscillator.

\section{Matrix representation of the operators}

If we consider each operator $T_i(x_i)$ as living in an $N-$ dimensional Hilbert space, then the functions
\begin{equation}\label{equ.hunt4ni}
\varphi _r (x_i ) = \sqrt {\frac{1}{L}} \cos \left[ {\frac{\pi }{2}\sin^2 \left(\frac{{r\pi }}{2}\right) - \frac{{(r+1)\pi x_i }}{2L}} \right],\quad r=0,1,2,\ldots,N-1,
\end{equation}
constitute a set of basis vectors of this $N-$ dimensional Hilbert space since they are the non-degenerate, mutually orthogonal and normalized eigenstates of the Hermitian operator $T_i$, with corresponding eigenvalues

\begin{equation}\label{equ.a9mebup}
\varepsilon_r=(r+1)^2\varepsilon,\quad r=0,1,2,\ldots,N-1,\mbox{ where } \varepsilon=\frac{\pi^2\hbar^2}{8mL^2}\,.
\end{equation}
Thus the operator $T(\vec x)$ lives in an $N^d$~dimensional Hilbert space whose basis vectors can be taken as the direct product vectors
\begin{equation}\label{equ.ble8tey}
\psi _s (\vec x)=\psi _s (x_1 ,x_2 , \ldots ,x_d ) = \prod_{i = 1}^d {\varphi _{s_i } (x_i )},\quad s=0,1,2,\ldots,N^d-1\,,
\end{equation}
where
\begin{equation}\label{equ.u88h55k}
s_i  = \left\lfloor {\frac{s}{{N^{d - i} }}} \right\rfloor \bmod N,\quad i = 1,2, \ldots ,d\,,
\end{equation}

where $\lfloor{q}\rfloor$, the {\em floor} of $q$, is the nearest integer not greater than $q$.

\bigskip

Thus each state $\psi _s (\vec x )$ is uniquely characterized or labelled by a vector \mbox{$\vec s=\left(s_1,s_2,\ldots,s_d\right)$} such that $s_i\in[0,1,2,\dots,N-1]$.

\bigskip

Denoting the eigenvalues of $T(\vec x)$ by $e_s$, we have
\[
e_s  = \sum_{i = 1}^d {\varepsilon _{s_i } }=\varepsilon\sum_{i = 1}^d{(s_i+1)^2},\quad s=0,1,2,\ldots,N^d-1\,, 
\]
with $s_i$ as defined in~\eqref{equ.u88h55k} and $\varepsilon$ as given in~\eqref{equ.a9mebup}.

\bigskip

Since the cho Hamiltonian $H(\vec x)$ lives in the same Hilbert space as $T(\vec x)$, the complete set of functions $\{\psi_s\}$, with $\psi_s$ as given in~\eqref{equ.ble8tey}, will be used as the basis vectors for the matrix representation of $H$.

\bigskip

Thus, for $s=0,1,2,\dots,N^d-1$ and $t=0,1,2,\dots,N^d-1$ and with $s_i$ and $t_i$ as given in~\eqref{equ.u88h55k}, the $N^{2d}$ matrix elements of $H$ are given by

\begin{equation}\label{equ.umfy2om}
\begin{split}
H_{st}  &= \left\langle {\psi _s } \right|H\left| {\psi _t } \right\rangle\\
 &= \sum_{i = 1}^d {\left\langle {\psi _s } \right|H_i \left| {\psi _t } \right\rangle }\\ 
 &= \sum_{i = 1}^d {\left\{ {\prod_{j = 0}^{d - 1} {\left\langle {\varphi _{s_{d - j} } (x_{d - j} )} \right|} H_i \prod_{j = 1}^d {\left| {\varphi _{t_j } (x_j )} \right\rangle } } \right\}}\\ 
 &= \sum_{i = 1}^d {\left\{ {\left( {\prod_{\scriptstyle j = 1 \hfill \atop 
  \scriptstyle j \ne i \hfill}^d {\delta _{s_j t_j } } } \right)\left( {\left\langle {\varphi _{s_i } } \right|H_i \left| {\varphi _{t_i } } \right\rangle } \right)} \right\}}\\ 
 &= \sum_{i = 1}^d {c_{i_{st} } H_{i_{s_i t_i } } }\,, 
\end{split}
\end{equation}

where we have introduced a \mbox{$d-$dimensional} vector $\vec c $ whose components are $N^d\times N^d$ symmetric binary matrices, $c_i$ with elements given by
\begin{equation}\label{equ.zjae2nl}
c_{i_{st} }  = \prod_{\scriptstyle j = 1 \hfill \atop 
  \scriptstyle j \ne i \hfill}^d {\delta _{s_i t_i } }\,,
\end{equation}
so that $ c_{i_{st} }=1 $ if either the two vectors $\vec s$ and $\vec t$ are one and the same vector, $\vec s=\vec t$, or they differ only at the $i^{th}$ component, otherwise $ c_{i_{st} } = 0$.

\bigskip

We note that
\begin{equation}\label{equ.h54j74c}
\delta _{s_i t_i } c_{i_{st} }  = \delta _{st}\,.
\end{equation}

\bigskip

In \eqref{equ.umfy2om}, $H_i,\;i=1,2,\dots d$ are $N^d\times N^d$, symmetric matrices with elements 
\begin{equation}\label{equ.dpldi3p}
\begin{split}
H_{i_{s_i t_i } }  &= \left\langle {\varphi _{s_i } } \right|H_i \left| {\varphi _{t_i } } \right\rangle\\
&= \left\langle {\varphi _{s_i } } \right|T_i \left| {\varphi _{t_i } } \right\rangle  + \left\langle {\varphi _{s_i } } \right|V_i \left| {\varphi _{t_i } } \right\rangle\\
&= T_{i_{_{s_i t_i } } }  + V_{i_{_{s_i t_i } } }\,,
\end{split}
\end{equation}
so that \eqref{equ.umfy2om} can now be written as $H_{st}=T_{st}+V_{st}$ with
\begin{equation}\label{equ.z9pe2a3}
T_{st}=\sum_{i=1}^d{c_{i_{st}}T_{i_{_{s_i t_i } } }}\,,\quad V_{st}=\sum_{i=1}^d{c_{i_{st}}V_{i_{_{s_i t_i } } }}
\end{equation}

\bigskip
 
We introduce yet another \mbox{$d-$dimensional} vector, $\vec\alpha$, whose components, $\alpha_i$, are $N^d\times N^d$ symmetric binary matrices, in terms of which the $c_i$ matrices may also be expressed. The $\alpha_i$ matrices are defined through their elements by $\alpha _{i_{st} } = \delta _{s_i t_i }$.

\subsection{Properties of the auxilliary matrices $c_i$ and $\alpha_i$}
It is straightforward to verify the following property for the $\alpha_i$ matrices:
\begin{equation}\label{equ.zzwhk4b}
\alpha _i \alpha _j  =\alpha _j \alpha _i= N^{d - 1} \delta _{ij} \alpha _i  + N^{d - 2} (1 - \delta _{ij} )J_{N^d }\,, 
\end{equation}
where
\[
J_{N^d }  = \left( {\begin{array}{*{20}c}
   1 & 1 &  \vdots  & 1  \\
   1 & 1 &  \vdots  & 1  \\
    \vdots  &  \vdots  &  \vdots  &  \vdots   \\
   1 & 1 &  \vdots  & 1  \\
\end{array}} \right)
\]
is the $N^d\times N^d$ {\it all-ones} matrix. The $\alpha_i$ matrices are singular and have trace equal to $N^d$. The eigenvalues of $\alpha_i$ are $N^{d-1}$ repeated $N$ times and $0$ repeated $N^d-N$ times. Finally using multinomial expansion theorem and \eqref{equ.zzwhk4b}, it is readily established that the matrix $\alpha=\sum_{i=1}^d\alpha_i$  satisfies
\[
\alpha^2 = N^{d - 1} \alpha  + N^{d - 2} d(d - 1)J_{N^d }\,.
\]
From~\eqref{equ.zjae2nl} it follows that
\begin{equation}\label{equ.b189fh0}
c_{i_{st} }  = \delta _{st}  + (1 - \alpha _{i_{st} } )\,\delta _{\alpha _{st} ,\,d - 1}
\end{equation}
and
\[
c_{st}=\sum_{i = 1}^d {c_{i_{st} } }  = \delta _{st}d  + \delta _{\alpha _{st} ,\,d - 1}\,.
\]
Explicitly
\[
c_{i_{st} }  = \left\{ {\begin{array}{*{20}c}
   \cos ^2 \left( {\alpha _{i_{st} }{\pi  \mathord{\left/
 {\vphantom {\pi  2}} \right.
 \kern-\nulldelimiterspace} 2}} \right) & \mbox{ if }{\alpha _{st}  = d - 1}  \\
 &\\

   0 & \mbox{ if }{\alpha _{st}  < d - 1}  \\
   &\\

   1 & \mbox{ if }{s = t}  \\
\end{array}} \right.
\]
and
\[
c_{st}  = \left\{ {\begin{array}{*{20}c}
   0 & \mbox{ if }{\alpha _{st}  < d - 1}  \\
   &\\

   1 & \mbox{ if }{\alpha _{st}  = d - 1}  \\
   &\\

   d & \mbox{ if }{s = t}  \\
\end{array}} \right.\,.
\]
From the definition of the $c_i$ matrices the following further properties are evident:
\begin{enumerate}
\item $c_i^n=N^{n-1}c_i$, for $n\in\mathbb{Z^+}$.
\item The eigenvalues of $c_i$ are $0$ and $N$, each being $N^{d-1}-$fold degenerate.
\item The $c_i$ matrices are singular and have trace $N^d$.
\end{enumerate}

\subsection{Representation for $T$}
Since $T_{i_{_{s_i t_i } } }=\left\langle {\varphi _{s_i } } \right|T_i \left| {\varphi _{t_i } } \right\rangle$, from~\eqref{equ.z9pe2a3} we have
\begin{equation}
\begin{split}
T_{st}&=\sum_{i=1}^dc_{i_{st}}\varepsilon(s_i+1)^2\delta_{s_it_i}\\
&=\varepsilon \delta _{st} \sum_{i = 1}^d {\left( {s_i  + 1} \right)^2 }\,, 
\end{split}
\end{equation}
where we have used \eqref{equ.hunt4ni}, \eqref{equ.a9mebup} and \eqref{equ.h54j74c}.
\subsection{Representation for $V$}
Since $V_{i_{_{s_i t_i } } }=\left\langle {\varphi _{s_i } } \right|V_i \left| {\varphi _{t_i } } \right\rangle$, performing the indicated integrations, with $\varphi _{s_i }$ and $\varphi _{t_i }$ as given in~\eqref{equ.hunt4ni} and noting that the only non-zero matrix elements of $V_i$ are those for which $s_i$ and $t_i$ are of the same parity, we obtain
\[
\begin{split}
V_{i_{s_i t_i } } &= \frac{{\lambda ^2 \varepsilon }}{2}\left[ {\frac{{\left( {1 - \delta _{s_i t_i } } \right)}}{{\left( {s_i  - t_i } \right)^2  + \delta _{s_i t_i } }} - \frac{{\left( {1 - \delta _{s_i t_i } } \right)}}{{\left( {s_i  + t_i  + 2} \right)^2 }}} \right]\cos ^2 \left[ {\left( {s_i  - t_i } \right)\frac{\pi }{2}} \right]\\
&\\
&\qquad+ \frac{{\lambda ^2 \varepsilon }}{{8}}\delta _{s_i t_i } \left[ {\frac{\pi ^2}{6}  - \frac{1}{{\left( {s_i  + 1} \right)^2 }}} \right]\,,
\end{split}
\]
where $\lambda=\omega\hbar/\varepsilon=\varepsilon_\omega/\varepsilon$.

\bigskip

Substituting for $V_{i_{s_i t_i } }$ in the second of \eqref{equ.z9pe2a3} and using \eqref{equ.h54j74c} and \eqref{equ.b189fh0}, we obtain
\[
\begin{split}
V_{st}  &= \frac{{\lambda ^2 \varepsilon }}{{8}}\delta _{st} \left[ {\frac{\pi ^2 d}{6} - \sum_{i = 1}^d {\frac{1}{{\left( {s_i  + 1} \right)^2 }}} } \right]\\
&\\
&\qquad+ \frac{{\lambda ^2 \varepsilon }}{2}\delta _{\alpha _{st} ,d - 1} \sum_{i = 1}^d {\left\{ {\left[ {\frac{{\left( {1 - \delta _{s_i t_i } } \right)}}{{\left( {s_i  - t_i } \right)^2  + \delta _{s_i t_i } }} - \frac{{\left( {1 - \delta _{s_i t_i } } \right)}}{{\left( {s_i  + t_i  + 2} \right)^2 }}} \right]\cos ^2 \left[ {\left( {s_i  - t_i } \right)\frac{\pi }{2}} \right]} \right\}}\,.
\end{split}
\]
We therefore see that off-diagonal survival of $V_{st}$ is possible (but not guarranteed due to the presence of the $\cos^2$ term) only if $\alpha _{st}=d - 1$, that is only if there exists a $k\in \left[1, d\right]$ such that $s_i=t_i$ if $i\ne k$ but $s_k\ne t_k$, so that the vectors $\vec r$ and $\vec s$ differ {\em only} at the $k^{th}$ entry.

\bigskip

Thus,
\[
\begin{split}
V_{st}  &= \frac{{\lambda ^2 \varepsilon }}{{8}}\delta _{st} \left[ {\frac{\pi ^2 d}{6} - \sum_{i = 1}^d {\frac{1}{{\left( {s_i  + 1} \right)^2 }}} } \right]\\
&\\
&\qquad+\frac{{\lambda ^2 \varepsilon }}{2}\delta _{\alpha _{st} ,d - 1} \left[ {\frac{1}{{\left( {s_k  - t_k } \right)^2}} - \frac{1}{{\left( {s_k  + t_k  + 2} \right)^2 }}} \right]\cos ^2 \left[ {\left( {s_k  - t_k } \right)\frac{\pi }{2}} \right]\,,
\end{split}
\]
where
\[
k = \sum_{j = 1}^d {j\left( {1 - \delta _{s_j t_j } } \right)}  = \sum_{j = 1}^d {j\left( {1 - \alpha _{j_{st} } } \right)}\,. 
\]
\subsection{Representation for $H$}

Adding the matrix elements $T_{st}$ and $V_{st}$ we find that the matrix elements for the Hamiltonian of the \mbox{$d-$dimensional} oscillator, with the direct product of eigenstates of the particle in a one dimensional box as basis, are given by
\[
\begin{split}
H_{st}  &= \varepsilon \delta _{st} \sum_{i = 1}^d {\left( {s_i  + 1} \right)^2 }  + \frac{{\lambda ^2 \varepsilon }}{{8}}\delta _{st} \left[ {\frac{\pi ^2 d}{6} - \sum_{i = 1}^d {\frac{1}{{\left( {s_i  + 1} \right)^2 }}} } \right]\\
&\\
&\qquad+\frac{{\lambda ^2 \varepsilon }}{2}\delta _{\alpha _{st} ,d - 1} \left[ {\frac{1}{{\left( {s_k  - t_k } \right)^2}} - \frac{1}{{\left( {s_k  + t_k  + 2} \right)^2 }}} \right]\cos ^2 \left[ {\left( {s_k  - t_k } \right)\frac{\pi }{2}} \right]\,,
\end{split}
\]
where
\[
k = \sum_{j = 1}^d {j\left( {1 - \delta _{s_j t_j } } \right)}  = \sum_{j = 1}^d {j\left( {1 - \alpha _{j_{st} } } \right)}\,. 
\]

\section{Application: Approximate analytic \mbox{expression} for the energy spectrum of the \mbox{$1-$dimensional} cho}
Based on our discussion in the previous sections culminating in the derivation of the explicit matrix elements of the \mbox{$d-$dimensional} confined harmonic oscillator, it is now straightforward, in principle, to find the eigenvalues of the oscillator. In practice however, the quantization remains a formidable task because of the exponential growth of the size of the Hamiltonian matrix with $d$. However, since the matrix elements are available in explicit form, they can be gainfully employed, for example in perturbation calculations, to obtain approximate analytical results.

\bigskip

It is our aim in the remaining part of this paper to derive an approximate analytic expression for the energy spectrum of the \mbox{$1-$dimensional} confined harmonic oscillator. We will treat the potential energy of the confined oscillator as a perturbation of the kinetic energy term, the latter being the exactly solvable particle in a box Hamiltonian, with the non-degenerate eigenstates given in~\eqref{equ.hunt4ni}. Results from perturbation calculations, in the one dimensional case, can also be found in references~\cite{baijal}, \cite{gueorguiev} and~\cite{padnos}.

\subsection*{Energy spectrum of the \mbox{$1-$dimensional} cho}
For the discussion of the confined harmonic oscillator in one dimension, it is convenient to drop the subscripts on $s$ and $t$. Also we shall refer to $H_i$, $V_i$, $T_i$ and $x_i$ simply as $H$, $V$, $T$ and $x$ respectively. The eigenvalue problem is therefore
\[
H\left| {E_r } \right\rangle  = E_r \left| {E_r } \right\rangle ,\quad r = 0,1,2, \ldots ,N-1\,.
\]
Since the eigenstates of $T(x)$ are known, being the $\varphi _r (x )$ of~\eqref{equ.hunt4ni}, it is convenient, for small values of the classical oscillator frequency, $\omega$, to treat $V(x)$ as a perturbation of the exactly solvable particle in a box Hamiltonian, $T(x)$, with $\omega^2$ as the perturbation parameter. 

\bigskip

By noting that \mbox{$\delta _{\alpha _{st} ,d - 1}  = \delta _{\alpha _{st} ,0}  = 1 - \delta _{st} $} we have
\begin{equation}\label{equ.jevxxpj}
\begin{split}
V_{st}  &= \frac{{\lambda ^2 \varepsilon }}{{8}}\delta _{st} \left[ {\frac{\pi ^2}{6} - {\frac{1}{{\left( {s  + 1} \right)^2 }}} } \right]\\
&\\
&\qquad+\frac{{\lambda ^2 \varepsilon }}{2}\left[ {\frac{{\left( {1 - \delta _{s t } } \right)}}{{\left( {s  - t } \right)^2  + \delta _{s t } }} - \frac{{\left( {1 - \delta _{s t } } \right)}}{{\left( {s  + t  + 2} \right)^2 }}} \right]\cos ^2 \left[ {\left( {s  - t } \right)\frac{\pi }{2}} \right]
\end{split}
\end{equation}
and
\[
\begin{split}
H_{st}  &= \varepsilon \delta _{st} {\left( {s  + 1} \right)^2 }  + \frac{{\lambda ^2 \varepsilon }}{{8}}\delta _{st} \left[ {\frac{\pi ^2}{6} - {\frac{1}{{\left( {s  + 1} \right)^2 }}} } \right]\\
&\\
&\qquad+\frac{{\lambda ^2 \varepsilon }}{2}\left[ {\frac{{\left( {1 - \delta _{s t } } \right)}}{{\left( {s  - t } \right)^2  + \delta _{s t } }} - \frac{{\left( {1 - \delta _{s t } } \right)}}{{\left( {s  + t  + 2} \right)^2 }}} \right]\cos ^2 \left[ {\left( {s  - t } \right)\frac{\pi }{2}} \right]\,.
\end{split}
\]

Since the states $\left| {E_r } \right\rangle$ are non-degenerate, $E_r$ can be approximated, using standard perturbation theory, as
\[
E_r  \approx E_r^{(0)}  + E_r^{(1)}  + E_r^{(2)}  + E_r^{(3)}~\,. 
\]
The zeroth order correction to the energy of the one dimensional confined harmonic oscillator, $ E_r^{(0)}$, being the energy of the one dimensional particle in a box and the first order correction, $ E_r^{(1)}$, being the expectation value of the perturbation $V(x)$, in each state $\varphi_r(x)$, are straightforward to calculate: 
\[
\begin{split}
E_r^{(0)}&=\varepsilon_r=\varepsilon(r+1)^2,\quad\mbox{from~\eqref{equ.a9mebup}}\\
&\mbox{and}\\
E_r^{(1)}=V_{rr}&=\frac{{\lambda ^2 \varepsilon }}{{8}}\left[ {\frac{\pi ^2}{6} - {\frac{1}{{\left( {r  + 1} \right)^2 }}} } \right],\quad\mbox{from~\eqref{equ.jevxxpj}}\\
&\\
&=\frac{{\lambda ^2 \varepsilon }}{{8}}\left[ {\zeta (2) - {\frac{1}{{\left( {r  + 1} \right)^2 }}} } \right]\,,
\end{split}
\]
where $\zeta(m)$ is the Riemann zeta function defined by
\[
\zeta (m) = \sum_{r = 1}^\infty  {\frac{1}{{r^m }}}\,. 
\]
The second order correction to the energy of the one dimensional confined harmonic oscillator, $ E_r^{(2)}$, is given by
\begin{equation}\label{equ.kzjql8y}
E_r^{(2)}  = \sum_{\scriptstyle s = 0 \hfill \atop 
  \scriptstyle s \ne r \hfill}^\infty {\frac{{V_{rs} V_{sr} }}{{\varepsilon _{rs} }}}  = \sum_{s = 0}^{r - 1} {\frac{{V_{rs} V_{sr} }}{{\varepsilon _{rs} }}}  + \sum_{s = r + 1}^\infty {\frac{{V_{rs} V_{sr} }}{{\varepsilon _{rs} }}}\,,
\end{equation}
where
\begin{equation}\label{equ.izptbkw}
\varepsilon _{rs}  = \varepsilon _r  - \varepsilon _s  = \varepsilon (r + s + 2)(r - s)\,,
\end{equation}
so that
\begin{equation}\label{equ.cxdserf}
\begin{split}
\frac{{2\varepsilon (r + 1)}}{{\varepsilon _{rs} }} &= \frac{1}{{(r + s + 2)}} + \frac{1}{{(r - s)}}\\
&\mbox{and}\\
\frac{{2\varepsilon (s + 1)}}{{\varepsilon _{rs} }} &=  - \frac{1}{{(r + s + 2)}} + \frac{1}{{(r - s)}}\,.
\end{split}
\end{equation}
Since $V$ is a real symmetric matrix, \eqref{equ.kzjql8y} is simply
\begin{equation}\label{equ.avhr7jt}
E_r^{(2)} = \sum_{s = 0}^{r - 1} {\frac{{V_{rs}^2 }}{{\varepsilon _{rs} }}}  + \sum_{s = r + 1}^\infty {\frac{{V_{rs}^2 }}{{\varepsilon _{rs} }}}\,.
\end{equation}
We note that the matrix elements occuring in~\eqref{equ.avhr7jt} are necessarily \mbox{off-diagonal} ($s\ne r$). Furthermore the only surviving elements $V_{rs}$, according to~\eqref{equ.jevxxpj}, are those for which $r$ and $s$ are both odd or both even. It therefore follows from~\eqref{equ.jevxxpj} that
\[
\begin{split}
V_{rs}  &= \frac{{\lambda ^2 \varepsilon }}{2}\left[ {\frac{1}{{(r - s)^2 }} - \frac{1}{{(r + s + 2)^2 }}} \right]\cos ^2 \left[ {\left( {s  - t } \right)\frac{\pi }{2}} \right]\\
&= \frac{{\lambda ^2 \varepsilon }}{2}\left( {\frac{1}{{(r - s) }} - \frac{1}{{(r + s + 2) }}} \right)\left( {\frac{1}{{(r - s) }} + \frac{1}{{(r + s + 2) }}} \right)\cos ^2 \left[ {\left( {r  - s } \right)\frac{\pi }{2}} \right]\,,
\end{split}
\]
and using~\eqref{equ.cxdserf} we have
\begin{equation}\label{equ.ifto5d5}
V_{rs}  = 2\lambda ^2 \varepsilon ^3 \frac{{(r + 1)(s + 1)}}{{\varepsilon _{rs}^2 }}\cos ^2 \left[ {\left( {r  - s } \right)\frac{\pi }{2}} \right]\,,\quad s\ne r\,.
\end{equation}
From~\eqref{equ.izptbkw} and \eqref{equ.ifto5d5} and noting that
\[
\cos ^4 \left[ {\left( {r  - s } \right)\frac{\pi }{2}} \right]\equiv\cos ^2 \left[ {\left( {r  - s } \right)\frac{\pi }{2}} \right]\,,
\]
we have
\[
\frac{{V_{rs}^2 }}{{\varepsilon _{rs} }} = 4\lambda ^4 \varepsilon \frac{{(r + 1)^2 (s + 1)^2 }}{{(r - s)^5 (r + s + 2)^5 }}\cos ^2 \left[ {\left( {r  - s } \right)\frac{\pi }{2}} \right]
\]
and thus~\eqref{equ.avhr7jt} now becomes
\begin{equation}\label{equ.whm0cis}
\begin{split}
E_r^{(2)}  &= 4\lambda ^4 \varepsilon (r + 1)^2 \sum\limits_{s = 0}^{r - 1} {\frac{{(s + 1)^2 }}{{(r - s)^5 (r + s + 2)^5 }}}\cos ^2 \left[ {\left( {r  - s } \right)\frac{\pi }{2}} \right]\\
&\qquad+ 4\lambda ^4 \varepsilon (r + 1)^2 \sum\limits_{s = r + 1}^\infty  {\frac{{(s + 1)^2 }}{{(r - s)^5 (r + s + 2)^5 }}}\cos ^2 \left[ {\left( {r  - s } \right)\frac{\pi }{2}} \right]\,.
\end{split}
\end{equation}
Classifying the energy corrections in~\eqref{equ.whm0cis} by parity of $r$ we have
\begin{equation}\label{equ.nrc1wbz}
E_q^{(2)}  = 4\lambda ^4 \varepsilon (q + 1)^2 \left( {A_q  + B_q } \right)\,,
\end{equation}
with
\[
\begin{split}
A_q  &= \sum\limits_{s = 0}^{q - 1} {\frac{{(s + 1)^2 }}{{(q - s)^5 (q + s + 2)^5 }}}\cos ^2 \left[ {\left( {q  - s } \right)\frac{\pi }{2}} \right]\\
&\text{and}\\
B_q  &= \sum\limits_{s = q + 1}^\infty  {\frac{{(s + 1)^2 }}{{(q - s)^5 (q + s + 2)^5 }}}\cos ^2 \left[ {\left( {q  - s } \right)\frac{\pi }{2}} \right]\,,
\end{split}
\]
where $q=2r$ for even levels and $q=2r+1$ for odd levels.

\bigskip
Choosing
\[
f_s  = \frac{{(s + 1)^2 }}{{(2r - s)^5 (2r + s + 2)^5 }}\cos ^2 \left[ {\left( {2r - s} \right)\frac{\pi }{2}} \right]
\]
in the following summation identity (see section~2.11 of~\cite{gould} for more general formulas)
\begin{equation}\label{equ.e5tgmvi}
\sum\limits_{s = 0}^M {f_s }  = \sum\limits_{s = 0}^{(M - (M\!\!\!\!\mod 2) )/2} {f_{2s} }  + \sum\limits_{s = 0}^{(M + (M\!\!\!\!\mod 2) )/2-1} {f_{2s + 1} }\,,
\end{equation}
allows us to write (noting that $f_{2s+1}=0$ with the present choice of $f_s$)
\begin{equation}\label{equ.y1qgeq0}
\begin{split}
A_{2r}&= \sum\limits_{s = 0}^{2r - 1} {\frac{{(s + 1)^2 }}{{(2r - s)^5 (2r + s + 2)^5 }}\cos ^2 \left[ {\left( {2r - s} \right)\frac{\pi }{2}} \right]}\\
&= \sum\limits_{s = 0}^{r - 1} {\frac{{(2s + 1)^2 }}{{(2r - 2s)^5 (2r + 2s + 2)^5 }}}\\
&= \frac{1}{{2^{10} }}\sum\limits_{s = 0}^{r - 1} {\frac{{(2s + 1)^2 }}{{(r - s)^5 (r + s + 1)^5 }}}
\end{split}
\end{equation}
and
\begin{equation}\label{equ.obkvl1s}
B_{2r}  = \frac{1}{{2^{10} }}\sum\limits_{s = r + 1}^\infty  {\frac{{(2s + 1)^2 }}{{(r - s)^5 (r + s + 1)^5 }}}\,.
\end{equation}
Similarly, taking identity~\eqref{equ.e5tgmvi} into consideration, we have
\begin{equation}\label{equ.n03mwer}
\begin{split}
A_{2r + 1}  &= \sum\limits_{s = 0}^{2r} {\frac{{(s + 1)^2 }}{{(2r - s + 1)^5 (2r + s + 3)^5 }}\cos ^2 \left[ {\left( {2r - s + 1} \right)\frac{\pi }{2}} \right]}\\
&= \sum\limits_{s = 0}^{r - 1} {\frac{{(2s + 2)^2 }}{{(2r - 2s)^5 (2r + 2s + 4)^5 }}}\\
&= \frac{4}{{2^{10} }}\sum\limits_{s = 0}^{r - 1} {\frac{{(s + 1)^2 }}{{(r - s)^5 (r + s + 2)^5 }}}
\end{split}
\end{equation}
and
\begin{equation}\label{equ.gvbcbjv}
B_{2r + 1}  = \frac{4}{{2^{10} }}\sum\limits_{s = r + 1}^\infty  {\frac{{(s + 1)^2 }}{{(r - s)^5 (r + s + 2)^5 }}}\,.
\end{equation}
We note that the above results can be combined into
\[
A_q  = \sum\limits_{s = 0}^{r - 1} {\frac{{(2s + 1 + q\bmod 2)^2 }}{{(2r - 2s)^5 (q + 2s + 2 + q\bmod 2)^5 }}} 
\]
and
\[
B_q  = \sum\limits_{s = r + 1}^\infty  {\frac{{(2s + 1 + q\bmod 2)^2 }}{{(2r - 2s)^5 (q + 2s + 2 + q\bmod 2)^5 }}}\,, 
\]
where $q=2r$ or $q=2r+1$.

\bigskip

The finite sums $A_{2r}$ and $A_{2r+1}$ as given in~\eqref{equ.y1qgeq0} and~\eqref{equ.n03mwer} as well as the infinite sums $B_{2r}$ and $B_{2r+1}$ as given in~\eqref{equ.obkvl1s} and~\eqref{equ.gvbcbjv} are expressible in closed form, in terms of the well-studied polygamma functions; a computer algebra system, such as Waterloo Maple, comes in handy for this purpose. Putting the results together in~\eqref{equ.nrc1wbz}, the final result is (see the Appendix for the Maple code)
\[
E_r^{(2)}  = \frac{{\lambda ^4 \varepsilon }}{{128}}\left( {\frac{{\zeta (4)}}{{(r + 1)^2 }} - \frac{{5\zeta (2)}}{{(r + 1)^4 }} + \frac{7}{{(r + 1)^6 }}} \right)\,,\quad r=0,1,2,\ldots
\]
We remark that an equivalent result to ours, for $E_r^{(2)} $, can also be found in reference~\cite{baijal}. The sum was however left unevaluated in that paper.

\bigskip

In standard non-degenerate perturbation theory, the third order correction to the energy of the one dimensional confined harmonic oscillator, $ E_r^{(3)}$, is given by
\[
E_r^{(3)}  = \sum_{\scriptstyle s = 0 \hfill \atop 
  \scriptstyle s \ne r \hfill}^N {\sum_{\scriptstyle t = 0 \hfill \atop 
  \scriptstyle t \ne r \hfill}^N {\frac{{V_{rs} V_{st} V_{tr} }}{{\varepsilon _{rs} \varepsilon _{rt} }}} }  - V_{rr} \sum_{\scriptstyle s = 0 \hfill \atop 
  \scriptstyle s \ne r \hfill}^N {\frac{{V_{rs} V_{sr} }}{{\varepsilon _{rs}^2 }}}\,. 
\]
Working exactly as in computing the second order corrections, while taking note of the following summation identity
\[
\sum_{s = a}^N {\sum_{t = a}^N {f_{st} } }  = \sum_{s = a}^N {f_{ss} }  + \sum_{s = a}^{N - 1} {\sum_{t = s + 1}^N {\left( {f_{st}  + f_{ts} } \right)} }\,, 
\]
we find that $E_r^{(3)}$ is expressible in closed form, in terms of the polygamma functions. In the limit of $N\to\infty$ the result is
\[
E_r^{(3)}  = \frac{{\lambda ^6 \varepsilon }}{{{\rm 2048}}}\left( {\frac{{\zeta (6)}}{{(r + 1)^4 }} - \frac{{60\zeta (4)}}{{(r + 1)^6 }} + \frac{{186\zeta (2)}}{{(r + 1)^8 }} - \frac{{242}}{{(r + 1)^{10} }}} \right)\,.
\]

Thus, the energy corrections can be written as
\[
E_r^{(m)}  = \frac{{\lambda ^{2m} \varepsilon }}{{2^{4m - 1} }}\sum_{n = 0}^m {\frac{{( - 1)^n \zeta (2m - 2n)c_n^{(m)} }}{{\left( {(r + 1)^2 } \right)^{m + n - 1} }}}\,,\quad m=0,1,2,3\,, 
\]
where
\begin{equation}\label{equ.gjqci09}
\begin{split}
c_0^{(0)}  &=  - 1\,,\\
&\\
c_0^{(1)}  &= 1,\,c_1^{(1)}  =  - 2\,,\\
&\\
c_0^{(2)}  &= 1,\,c_1^{(2)}  = 5,\,c_2^{(2)}  =  - 14\\
\mbox{and}\\
c_0^{(3)}  &= 1,\,c_1^{(3)}  = 60,\,c_2^{(3)}  = 186,\,c_3^{(3)}  =  - 484\,.
\end{split}
\end{equation}
To the sixth order in the classical oscillator frequency, $\omega$, therefore, the one dimensional cho has the energy spectrum
\begin{equation}\label{equ.heoyd4m}
E_r  \approx \sum_{m = 0}^3 {E_r^{(m)} }  = \sum_{m = 0}^3 {\left\{ {\frac{{\lambda ^{2m} \varepsilon }}{{2^{4m - 1} }}\sum_{n = 0}^m {\frac{{( - 1)^n \zeta (2m - 2n)c_n^{(m)} }}{{\left( {(r + 1)^2 } \right)^{m + n - 1} }}} } \right\}}\,,
\end{equation}
with $c_n^{(m)}$ as given in~\eqref{equ.gjqci09}.

\bigskip

The form of~\eqref{equ.heoyd4m} allows to conjecture the existence of an exact formula for the energy spectrum of the one dimensional confined harmonic oscillator, in the form
\[
E_r  = \sum_{m = 0}^\infty  {E_r^{(m)} }  = \sum_{m = 0}^\infty  {\left\{ {\frac{{\lambda ^{2m} \varepsilon }}{{2^{4m - 1} }}\sum_{n = 0}^m {\frac{{( - 1)^n \zeta (2m - 2n)c_n^{(m)} }}{{\left( {(r + 1)^2 } \right)^{m + n - 1} }}} } \right\}}\,,\quad c_n^{(m)}\in\mathbb{Z}\backslash\{0\}\,. 
\]
\section{Summary}
We have derived an explicit matrix representation for the $d-$dimensional confined harmonic oscillator, using the eigenstates of the kinetic energy operator as basis vectors.

\bigskip

We showed that the Hamiltonian
\[
H(\vec x)=-\frac{\hbar^2}{2m}\sum_{i=1}^d{\frac{\partial^2}{\partial x_i^2}+\frac 12m\omega^2\sum_{i=1}^d{x_i^2}}\,,\quad |x_i|\le L\,,
\]
has the explicit $N^d\times N^d$ matrix representation
\[
\begin{split}
H_{st}  &= \varepsilon \delta _{st} \sum_{i = 1}^d {\left( {s_i  + 1} \right)^2 }  + \frac{{\lambda ^2 \varepsilon }}{{8}}\delta _{st} \left[ {\frac{\pi ^2 d}{6} - \sum_{i = 1}^d {\frac{1}{{\left( {s_i  + 1} \right)^2 }}} } \right]\\
&\\
&\qquad+\frac{{\lambda ^2 \varepsilon }}{2}\delta _{\alpha _{st} ,d - 1} \left[ {\frac{1}{{\left( {s_k  - t_k } \right)^2}} - \frac{1}{{\left( {s_k  + t_k  + 2} \right)^2 }}} \right]\cos ^2 \left[ {\left( {s_k  - t_k } \right)\frac{\pi }{2}} \right]\,,
\end{split}
\]
with \[\varepsilon=\frac{\pi^2\hbar^2}{8mL^2}, \lambda=\omega\hbar/\varepsilon=\varepsilon_\omega/\varepsilon, \alpha_{st}=\sum_{i=1}^d{{\alpha_i}_{st}}=\sum_{i=1}^d{\delta_{s_it_i}}\] and
\[
k = \sum_{j = 1}^d {j\left( {1 - \delta _{s_j t_j } } \right)}  = \sum_{j = 1}^d {j\left( {1 - \alpha _{j_{st} } } \right)}\,, 
\]
where $s,t=0,1,2,\ldots,N^d-1$ and
\[
s_i  = \left\lfloor {\frac{s}{{N^{d - i} }}} \right\rfloor \bmod N,\quad i = 1,2, \ldots ,d
\]
and
\[
t_i  = \left\lfloor {\frac{t}{{N^{d - i} }}} \right\rfloor \bmod N,\quad i = 1,2, \ldots ,d\,.
\]
In particular, for the one-dimensional confined harmonic oscillator, we have an $N\times N$ representation with the matrix elements given by
\[
\begin{split}
H_{st}  &= \varepsilon \delta _{st} {\left( {s  + 1} \right)^2 }  + \frac{{\lambda ^2 \varepsilon }}{{8}}\delta _{st} \left[ {\frac{\pi ^2}{6} - {\frac{1}{{\left( {s  + 1} \right)^2 }}} } \right]\\
&\\
&\qquad+\frac{{\lambda ^2 \varepsilon }}{2}\left[ {\frac{1}{{\left( {s  - t } \right)^2  + \delta _{s t } }} - \frac{1}{{\left( {s  + t  + 2} \right)^2 }}} \right]{\left( {1 - \delta _{s t } } \right)}\cos ^2 \left[ {\left( {s  - t } \right)\frac{\pi }{2}} \right]\,,
\end{split}
\]
for $s,t=0,1,2,\ldots,N-1$.

\bigskip

Finally, we derived the following approximate analytic expression for the energy spectrum of the \mbox{$1-$dimensional} cho, to the sixth order in the oscillator frequency $\omega$,  
\[
E_r  \approx \sum_{m = 0}^3 {E_r^{(m)} }  = \sum_{m = 0}^3 {\left\{ {\frac{{\lambda ^{2m} \varepsilon }}{{2^{4m - 1} }}\sum_{n = 0}^m {\frac{{( - 1)^n \zeta (2m - 2n)c_n^{(m)} }}{{\left( {(r + 1)^2 } \right)^{m + n - 1} }}} } \right\}}\,,\quad r=0,1,2,\ldots,N\,,
\]
with $c_n^{(m)}$ as given in~\eqref{equ.gjqci09}.

\section*{Appendix}
\subsection*{Maple code to determine $E_r^{(2)}$}
=================================================

\verb|>summand:=q->(2*s+1+modp(q,2))^2/(2*r-2*s)^5/(q+2*s+2+modp(q,2))^5;|
\[
summand: = q \to \frac{{(2s + 1 + {\rm modp}(q,2))^2 }}{{(2r - 2s)^5 (q + 2s + 2 + {\rm modp}(q,2))^5 }}
\]
 
\verb|>A2r:=sum(summand(2*r),s=0..r-1):|

\verb| # replace the last ":" with ";" to see the polygamma sums|

\verb|>B2r:=sum(summand(2*r),s=r+1..infinity):|

\verb|# replace ":" with ";" to see the polygamma sums|

\verb|>E2r:=expand(simplify(4*(2*r+1)^2*(A2r+B2r))):|

\verb|# we suppress the factor [lambda^4*epsilon]|

\verb|># collect terms of the same order in Pi|

\verb|E2r:=collect(%,Pi):|

\verb|>E2r:=factor(coeff(E2r,Pi^4))*Pi^4+factor(coeff(E2r,Pi^2))*Pi^2+op(3,E2r);|
\[
E2r:={\frac {{\pi }^{4}}{11520\, \left( 2\,r+1 \right) ^{2}}}-{\frac {5\,{
\pi }^{2}}{768\, \left( 2\,r+1 \right) ^{4}}}+{\frac {7}{128\, \left( 
2\,r+1 \right) ^{6}}}
\]

\verb|># we now include the lambda^4*epsilon|

\verb|E2r:=lambda^4*epsilon*E2r;|
\[
E2r:=\lambda^4\varepsilon \left({\frac {{\pi }^{4}}{11520\, \left( 2\,r+1 \right) ^{2}}}-{\frac {5\,{
\pi }^{2}}{768\, \left( 2\,r+1 \right) ^{4}}}+{\frac {7}{128\, \left( 
2\,r+1 \right) ^{6}}}\right)
\]
=================================================

\bigskip

Since $\zeta(2)=\pi^2/6$ and $\zeta(4)=\pi^4/90$ it follows that
\begin{equation}\label{equ.j4lup9c}
E_{2r}^{(2)}  = \frac{{\lambda ^4 \varepsilon }}{{128}}\left( {\frac{{\zeta (4)}}{{(2r + 1)^2 }} - \frac{{5\zeta (2)}}{{(2r + 1)^4 }} + \frac{7}{{(2r + 1)^6 }}} \right)\,.
\end{equation}

\bigskip

Similarly running the above code with $summand(2r+1)$ gives
\begin{equation}\label{equ.airhzmn}
E_{2r+1}^{(2)}  = \frac{{\lambda ^4 \varepsilon }}{{128}}\left( {\frac{{\zeta (4)}}{{(2r + 2)^2 }} - \frac{{5\zeta (2)}}{{(2r + 2)^4 }} + \frac{7}{{(2r + 2)^6 }}} \right)\,.
\end{equation}
From~\eqref{equ.j4lup9c} and \eqref{equ.airhzmn} we conclude that
\[
E_r^{(2)}  = \frac{{\lambda ^4 \varepsilon }}{{128}}\left( {\frac{{\zeta (4)}}{{(r + 1)^2 }} - \frac{{5\zeta (2)}}{{(r + 1)^4 }} + \frac{7}{{(r + 1)^6 }}} \right)\,.
\]

\end{document}